\documentclass{epl}

\def\zp {Z^{\pm}}
\def\zm {Z^{\mp}}
\def\Dzp {\Delta Z^{\pm}}
\def\Dzm {\Delta Z^{\mp}}

\def\dt {\partial_t}
\def\da {\partial_{\alpha}}

\def\daq {\partial_{\alpha}^2}
\def\da2p {\partial_{\alpha}^{2 \prime}}

\def\di {\partial_i}
\def\dip {\partial_i^{\prime}}

\def\bx {\mathbf{x}}

\usepackage{graphicx}
\usepackage{dcolumn}
\usepackage{bm}
\input{epsf}

\title{On the turbulent energy cascade in anisotropic Magnetohydrodynamic turbulence}

\author{V. Carbone\inst{1,2} et al.
}
\institute{
  \inst{1} Dipartimento di Fisica, Universit\`a della Calabria, Rende (CS), Italy \\ 
  \inst{2} Liquid Crystal Laboratory (CNR), Rende (CS), Italy 
}
\pacs{52.30.Cv}{Magnetohydrodynamics (including electron magnetohydrodynamics)}
\pacs{94.05.Lk}{Turbulence}
\pacs{96.50.Tf}{MHD waves, plasma waves, turbulence in interplanetary physics}

\begin{document}

\maketitle

\begin{abstract}

The problem of the occurrence of an energy cascade for Alfv\'enic turbulence in solar wind plasmas was hystorically addressed by using phenomenological arguments based to the weakness of nonlinear interactions and the anisotopy of the cascade in wave vectors space. Here, this paradox is reviewed through the formal derivation of a Yaglom relation from anisotropic Magnetohydrodynamic equation. The Yaglom relation involves a third-order moment calculated from velocity and magnetic fields and involving both Els\"asser vector fields, and is particularly useful to be used as far as spacecraft obervations of turbulence are concerned. 

\end{abstract}

\section{Introduction}

In a seminal paper, Dobrowolny, Mangeney and Veltri~\cite{DMV} (hereafter DMV), rised the question of the existence of an energy cascade in Magnetohydrodynamic (MHD) turbulence because the apparent contradiction of two competing observations within the solar wind turbulence by \textit{in situ} satellite measurements. In fact, since the oldest space flights, both a well defined turbulent spectrum, and strong correlations between velocity and magnetic field fluctuations have been observed~\cite{coleman,belcher} (for a modern review of turbulence in the solar wind cfr. Ref.~\cite{living}). The apparent contradiction between these observations can be immediately seen by introducing the Els\"asser variables

\[
\zp(\bx,t) = \mathbf{v} \pm \frac{\mathbf{B}}{\sqrt{4\pi \rho}}
\] 
where $\mathbf{v}_i$ and $\mathbf{B}_i$ represent the velocity and magnetic field respectively, while $\rho$ is the mass density. These quantities represent Alfv\'enic fluctuations propagating along the background magnetic field, in opposite directions. MHD equations can be immediately written in terms of these variables as

\begin{equation}
\dt \zp_i + \left(\zm_\alpha \partial_\alpha \right) \zp_i = -\di \pi + \lambda^{\pm} \daq Z^+_i + \lambda^{\mp} \daq Z^-_i
\label{mhd}
\end{equation}
where $\pi = P/\rho$ ($P$ being the the total pressure), $\dt$ represents time derivative while $\di$ represents derivative with respect to the spatial variable $x_i$. The kinematic viscosity $\nu$ and the magnetic diffusivity $\mu$ form the dissipative coefficients $\lambda^{\pm} = (\nu \pm \mu)/2$. The second term in equations~(\ref{mhd}) shows that nonlinear interactions only occur between opposite sign fluctuations. Since high correlations between velocity and magnetic fluctuations imply either $\zm_i = 0$ or $\zp_i = 0$, a turbulent energy cascade should be incompatible with the disappearence of one of the alfv\'enic fluctuations. The puzzle have been apparently solved by DMV~\cite{DMV}. In presence of a strong magnetic field, nonlinear interactions are slowed down by the transport of fluctuations (Alfv\'en effect). The usual Kolmogorov's phenomenlogy must then be modified in favor of the Iroshnikov-Kraichnan (IK). This yields to the fact that the energy transfer rates per unit mass for both pseudo-energies associated to alfv\'enic fluctuations must be of the same order, $\epsilon^+ \sim \epsilon^-$. More precisely, they must have the same scaling laws in the IK phenomenology~\cite{carbo93}. Thus, an initial small unbalance between alfv\'enic fluctuations is maintained during the cascade, eventually leading to both a turbulent spectrum, and high correlations~\cite{DMV}. This framework is referred to in the literature as Alfv\'enic turbulence. The above arguments have been criticized~\cite{critica1,critica2} on the basis of the fact that, at variance with the conjecture in Ref.~\cite{DMV}, both in closures equations~\cite{critica1} and in direct numerical simulations~\cite{critica2} the energy transfer rates are never the same. 
A different phenomenological argument has also been introduced~\cite{anisotropia}. In a strongly anisotropic medium as the solar wind, the energy cascade will eventually develop in the direction transverse to the background magnetic field. Using the hypothesis of a critical balance between the time of transport in the parallel direction and the eddy turnover time for turbulence in the transverse direction, a Kolmogorov's spectrum, rather than a Kraichnan's spectrum, is expected for transverse wavevectors. 

In homogeneous and isotropic fluid turbulence, the energy cascade is evidenced by the observation of a well defined relation between the third-order longitudinal structure function and the energy dissipation transfer rate, namely the well known $4/5$-Kolmogorov's law~\cite{frisch}. A similar relation have been derived for MHD, following the Yaglom law for passive scalars, in the framework of homogeneous and isotropic turbulence~\cite{pp}. This law has been recently observed in samples of ecliptic~\cite{smith} and polar~\cite{noi} solar wind, thus showing unambiguously that a turbulent cascade is active. These observations solve the apparent paradox rised in the DMV paper, confirming the presence of both strong correlations and turbulent cascade. However, the anisotropic nature of solar wind turbulence occasionally violates the conditions for the observation of the Yaglom law, namely of the cascade. In this letter we analyze the relevance of anisotropy conjecture in deriving the Yaglom's scaling law for MHD turbulence.


\section{The Yaglom law for anisotropic and isotropic MHD}

Consider the anisotropic MHD equations~(\ref{mhd}) written twice for Els\"asser variable $Z_i^{\pm}(x_i)$ at the point $x_i$, and for $Z_i^{\pm}(x_i+r_i)$ at the independent point $x_i^{\prime} = x_i + r_i$. By substraction, we obtain an equation for the differences $\Dzp_i = (Z_i^{\pm}) ^{\prime} -\zp_i$ (here and in the following ``primed'' variables are intended as calculated on the point $x_i^{\prime}$). Using the hypothesis of independence of points $x_i^{\prime}$ and $x_i$ with respect to derivatives, namely $\di (Z_j^{\pm})^{\prime} = \di^{\prime} Z_j^{\pm} = 0$ (where $\di^{\prime}$ represents derivative with respect to $x_i^{\prime}$), we get

\begin{equation}
\dt \Dzp_i + Z_\alpha^{\mp \prime} \partial_\alpha^{\prime} \Dzp_i = -(\dip + \di) \Delta P + (\partial_\alpha^{2 \prime} + \partial_\alpha^2) \left[\lambda^{\pm} \Delta Z^+_i + \lambda^{\mp} \Delta Z^-_i\right]
\label{prima}
\end{equation}
($\Delta P = \pi^{\prime} - \pi$). By adding and substracting the term $\zm_\alpha \partial_\alpha^{\prime} \Dzp_i$ to (\ref{prima}) we obtain

\begin{eqnarray}
\dt \Dzp_i + \Delta Z_\alpha^{\mp} \partial_\alpha^{\prime} \Dzp_i &+& \zm_\alpha (\partial_\alpha^{\prime} + \partial_\alpha) \Dzp_i = -(\dip + \di) \Delta P + \nonumber \\
&+& (\partial_\alpha^{2 \prime} + \partial_\alpha^2) \left[\lambda^{\pm} \Delta Z^+_i + \lambda^{\mp} \Delta Z^-_i\right]
\label{seconda}
\end{eqnarray}

We are seeking for an equation for the second-order correlation tensor $\langle \Dzp_i \Dzp_j \rangle$ related to pseudo-energies. In fact, in a more general approach one should look at a mixed tensor, namely $\langle \Dzp_i \Dzm_j \rangle$, taking into account not only both pseudo-energies  but also cross-helicity $\langle Z_i^{+} Z_j^{-} \rangle$ and $\langle Z_i^{-} Z_j^{+} \rangle$. However, using the DIA closure by Kraichnan, it is possible to show that these elements are in general poorly correlated~\cite{veltri}. Since we are interested in the energy cascade, we limit ourself to the most interesting equation that describes correlations about Alfv\'enic fluctuations of the same sign. To obtain the equations for pseudo-energies we multiply equations (\ref{seconda}) by $\Dzp_j$, then by averaging we get 

\begin{eqnarray}
\dt \langle \Dzp_i \Dzp_j \rangle &+& \langle \Delta Z_\alpha^{\mp} \partial_\alpha^{\prime} (\Dzp_i \Dzp_j) \rangle + \langle \zm_\alpha (\partial_\alpha^{\prime} + \partial_\alpha) (\Dzp_i \Dzp_j) \rangle = \nonumber \\ 
= &-& \langle \Dzp_j(\dip + \di) \Delta P + \Dzp_i (\partial_j^{\prime} + \partial_j) \Delta P \rangle 
+ \nonumber \\
&+& \lambda^{\pm} \langle \Dzp_j (\partial_\alpha^{2 \prime} + \partial_\alpha^2) \Delta Z^+_i \rangle + \lambda^{\pm} \langle \Dzp_i (\partial_\alpha^{2 \prime} + \partial_\alpha^2) \Delta Z^+_j \rangle + \nonumber \\
&+& \lambda^{\mp} \langle \Dzp_j (\partial_\alpha^{2 \prime} + \partial_\alpha^2) \Delta Z^-_i \rangle + \lambda^{\mp} \langle \Dzp_i (\partial_\alpha^{2 \prime} + \partial_\alpha^2) \Delta Z^-_j \rangle
\label{terza}
\end{eqnarray}

If we consider local homogeneity we have

\[
\partial_\alpha^{\prime} \equiv \frac{\partial}{\partial (x_\alpha + r_\alpha)} \simeq \frac{\partial}{\partial r_\alpha}
\]

\[
\partial_\alpha \equiv \frac{\partial}{\partial (x_\alpha^{\prime} - r_\alpha)} \simeq -\frac{\partial}{\partial r_\alpha}
\]
when applied to difference quantities, so that the nonlinear term, using incompressibility, becomes

\[
\langle \Dzm_\alpha \partial_\alpha^{\prime} (\Dzp_i \Dzp_j) \rangle = \frac{\partial}{\partial r_\alpha}
\langle \Dzm_\alpha (\Dzp_i \Dzp_j) \rangle
\]

Note that in eq.~(\ref{mhd}) kinematic viscosity are not assumed equal to magnetic diffusivity, and this generates a coupling between $\zp_i$ and $\zm_i$ not only in the nonlinear term but also in the dissipative term. We exclude these couplings by making here the usual symplifying assumption, that kinematic viscosity is equal to magnetic diffusivity, $\lambda^{\pm} = \lambda^{\mp} = \nu$. Then, by using the independence of derivatives with respect to both points and using the local homogeneity hypothesis, the dissipative term becomes

\[
\nu \langle (\partial_\alpha^{2 \prime} + \partial_\alpha^2) (\Dzp_i \Dzp_j) \rangle = 2 \nu \frac{\partial^2}{\partial r_\alpha} \langle \Dzp_i \Dzp_j \rangle - \frac{4}{3} \frac{\partial}{\partial r_\alpha} (\epsilon_{ij}^{\pm} r_\alpha)
\]
where we defined the average dissipation tensor

\begin{equation}
\epsilon_{ij}^{\pm} = \nu \langle (\partial_\alpha Z_i^{\pm}) (\partial_\alpha Z_j^{\pm}) \rangle
\label{dissipa}
\end{equation}
Using these equations in~(\ref{terza}), we finally obtain the equation 

\begin{eqnarray}
\dt \langle \Dzp_i \Dzp_j \rangle &+& \frac{\partial}{\partial r_\alpha}\langle \Delta Z_\alpha^{\mp} (\Dzp_i \Dzp_j) \rangle = \nonumber \\ 
&=& - \Lambda_{ij} - \Pi_{ij} + 2 \nu \frac{\partial^2}{\partial r_\alpha^2} \langle \Dzp_i \Dzp_j \rangle - \frac{4}{3} \frac{\partial}{\partial r_\alpha} (\epsilon_{ij}^{\pm} r_\alpha)
\label{anisotropa}
\end{eqnarray}
The first and second term on the r.h.s. of the last equation represent respectively a tensor related to large-scale inhomogeneities $\Lambda_{ij} = \langle \zm_\alpha (\partial_\alpha^{\prime} + \partial_\alpha) (\Dzp_i \Dzp_j) \rangle$, and the tensor related to the pressure term $\Pi_{ij} = \langle \Dzp_j(\dip + \di) \Delta P + \Dzp_i (\partial_j^{\prime} + \partial_j) \Delta P \rangle$. Equation~(\ref{anisotropa}) is an exact equation for anisotropic MHD equations that links the second-order complete tensor to the third-order mixed tensor through the average dissipation rate tensor. 

Using incompressibility and independence of derivatives with respect to both points, the first term on the r.h.s. can be written as $\Lambda_{ij} = (\partial_\alpha^{\prime} + \partial_\alpha) \langle \zm_\alpha (\Dzp_i \Dzp_j) \rangle$ which vanishes for a globally homogeneous situation, because in this case $\di \langle \rangle \equiv 0$. The pressure term is more complicated to be managed. Using independence of derivatives and local homogeneity we get 

\begin{eqnarray}
\langle \Dzp_i \partial_j\Delta P \rangle &=& \langle \partial_j[\Dzp_i \Delta P] - (\partial_j z^{\pm}_i) \Delta P \rangle = \\ \nonumber
&=& - \langle \partial_j^{\prime}[\Dzp_i \Delta P] \rangle - \langle (\partial_j z^{\pm}_i) \Delta P \rangle \nonumber
\end{eqnarray}
from which

\begin{equation}
\Pi_{ij} = \langle \left[\partial_j^{\prime} (z^{\pm}_i)^{\prime} - \partial_j z^{\pm}_i \right] \Delta P \rangle + \langle \left[\partial_i^{\prime} (z^{\pm}_j)^{\prime} - \partial_i z^{\pm}_j \right] \Delta P \rangle
\label{pressure}
\end{equation}
Then the diagonal terms of the tensor containing the pressure vanish. In fact summing over indices eq. (\ref{pressure}) yields $[\dip (z^{\pm}_{i})^{\prime} - \di z^{\pm}_i]$ which is zero for local homogeneity and incompressibility. This means that, assuming global homogeneity and incompressibility, the equation for the trace of tensor can be written as

\begin{equation}
\dt \langle |\Dzp_i|^2 \rangle + \frac{\partial}{\partial r_\alpha}\langle \Delta Z_\alpha^{\mp} |\Dzp_i|^2 \rangle = 2 \nu \frac{\partial^2}{\partial r_\alpha} \langle |\Dzp_i|^2 \rangle - \frac{4}{3} \frac{\partial}{\partial r_\alpha} (\epsilon^{\pm}_{ii} r_\alpha)
\label{traccia}
\end{equation}
This expression is valid even in the anisotropic case, that is fields depends on the vector $r_\alpha$. Moreover by considering only the trace, we ruled out the possibility to investigate anisotropies related to different orientations of vectors within the second-order moment. It is worthwhile to remark here that \textit{only} the diagonal elements of the dissipation rate tensor, namely $\epsilon_{ii}^{\pm}$ are positive defined, while in general the off-diagonal elements $\epsilon^{\pm}_{ij}$ can be in principle also negative. 
For a stationary state the equation (\ref{traccia}) can be written as the divergenceless condition of a quantity involving the third-order correlations and the dissipation rates

\begin{equation}
\frac{\partial}{\partial r_\alpha} \left[ \langle \Delta Z_\alpha^{\mp} |\Dzp_i|^2 \rangle - 2 \nu \frac{\partial}{\partial r_\alpha} \langle |\Dzp_i|^2 \rangle - \frac{4}{3} (\epsilon^{\pm}_{ii} r_\alpha) \right] = 0
\label{anisotropa1}
\end{equation}
from which we can obtain the Yaglom's relation by projecting equation (\ref{anisotropa1}) along the longitudinal $r_\alpha = r \mathbf{e}_r$ direction. This operation involves the assumption that the flow is locally isotropic, that is fields depends locally only on the separation $r$, so that 

\begin{equation}
\left(\frac{2}{r}+\frac{\partial}{\partial r}\right)\left[\langle \Delta Z_r^{\mp} |\Dzp_i|^2 \rangle - 2\nu \frac{\partial}{\partial r} \langle |\Dzp_i|^2 \rangle + \frac{4}{3} \epsilon^{\pm}_{ii} r \right] = 0
\label{equat}
\end{equation}
The only solution that is compatible with the absence of singularity in the limit $r \to 0$ is 

\begin{equation}
\langle \Delta Z_r^{\mp} |\Dzp_i|^2 \rangle = 2\nu \frac{\partial}{\partial r} \langle |\Dzp_i|^2 \rangle - \frac{4}{3} \epsilon^{\pm}_{ii} r
\label{yaglomdis}
\end{equation}
which reduces to the Yaglom's law for MHD turbulence as obtained by Politano and Pouquet \cite{pp} in the inertial range when $\nu \to 0$

\begin{equation}
\langle \Delta Z_r^{\mp} |\Dzp_i|^2 \rangle = - \frac{4}{3} \epsilon^{\pm}_{ii} r
\label{yaglom}
\end{equation}
Finally, in the fluid-like case where $z^{+}_i = z^{-}_i = v_i$ we obtain the usual Yaglom's law $\langle \Delta v_r |\Delta v_i|^2 \rangle = - 4/3 \left(\epsilon r \right)$ ($\epsilon$ being the usual dissipation rate) which immediately reduces to the Kolmogorov's law $\langle \Delta v_r^3 \rangle = - 4/5 \left(\epsilon r \right)$ in the isotropic case where $\langle \Delta v_r \Delta v_y^2 \rangle = \langle \Delta v_r \Delta v_z^2 \rangle = 1/3 \langle \Delta v_r^3 \rangle$ (assuming the separation $r$ along the streamwise direction $x$).

Even if eq. (\ref{yaglom}) remains formally valid only for isotropic MHD, this cannot completely solve the problem of the energy cascade in MHD. The tensor $\Pi_{ij}$ is zero only when local anisotropy is assumed. In fact by calculating the divergence with respect to the index $i$ of $\Pi_{ij}$, assuming independence of derivatives, we get

\begin{equation}
\partial_i \langle \Dzp_i(\partial_j P - \partial_j^{\prime} P^{\prime} \rangle = \langle \Dzp_i \partial_i \partial_j P \rangle = \partial_i \langle \Dzp_i \partial_j P \rangle = - \langle \partial_i^{\prime} \Dzp_i \partial_j P \rangle = 0
\label{equ2}
\end{equation}
By symmetry the divergence with respect to $j$ also vanishes, that is $\partial_{ij}^2 \Pi_{ij} = 0$. By using for $\Pi_{ij}$ the isotropic formula for a generic tensor \cite{monin} 

\[
\Pi_{ij}(\mathbf{r}) = \left[\Pi_{11}(r)-\Pi_{\alpha \alpha}(r)\right]\frac{r_i r_j}{r^3} + \Pi_{\alpha \alpha}(r) \delta_{ij}
\]
it can be easily shown that $\Pi_{ij} = 0$ by local isotropy \cite{monin}.

\section{Reduced MHD}

As a different approach, let us consider the Reduced MHD approximation (RMHD)~\cite{rmhd,rmhd1} which is valid under the hypothesis of a strong guide magnetic field. In this approximation, the dynamics along the parallel and perpendicular directions (with respect to the average magnetic field) are disentangled. Then we can distinguish between derivatives for perpendicular ($\partial/\partial x_\perp$) and parallel ($\partial/\partial x_\parallel$) coordinates. In terms of Els\"asser variables, the RMHD approximation reads~\cite{nigro}

\begin{equation}
\frac{\partial \zp_\perp}{\partial t} + \zm_\perp \frac{\partial \zp_\perp}{\partial x_\perp} \pm \frac{\partial \zp_\perp}{\partial x_\parallel} = - \frac{\partial \pi}{\partial x_\perp} + \lambda^{\pm} \frac{\partial^2 Z^+_\perp}{\partial x_{\perp}^2} + \lambda^{\mp} \frac{\partial^2 Z^-_\perp}{\partial x_{\perp}^2}
\label{rmhd}
\end{equation}
where $Z^{\pm}_{\perp}(x_\parallel,x_\perp)$ is the perpendicular component of the Els\"asser variables. From this equations, by performing the same calculations as before, and by defining the separations along the perpendicular $r_\perp$ and parallel $r_\parallel$ directions respectively, we obtain the following K\'arm\'an-Howarth relation for the stationary state

\begin{equation}
\langle \Dzm_{\perp} |\Dzp_\perp|^2 \rangle = 2 \nu \frac{\partial}{\partial r_\perp} \langle |\Dzp_\perp|^2 \rangle -2 \epsilon_{ii}^{\pm} r_\perp + \frac{2}{r_\perp} \int_0^{r_{\perp}} r_{\perp}^{\prime} \frac{\partial}{\partial r_\parallel} \langle |\Dzp_\perp|^2 \rangle dr_{\perp}^{\prime}
\label{yaglomrmhd}
\end{equation}
From this last equation it is evident that, in the limit of vanishing viscosity $\nu \to 0$, in general a Yaglom's relation cannot be derived. This can only be the case by assuming that the average pseudo-energies are almost constant along the parallel direction. So, if the third term on the r.h.s. is zero, we can derive a Yaglom's relation between the mixed third-order correlation term as a linear function of the transverse scale $r_\perp$

\begin{equation}
\langle \Dzm_{\perp} |\Dzp_i|^2 \rangle \sim -2 \epsilon_{ii}^{\pm} r_\perp
\label{yaglomrmhd1}
\end{equation}
Even if this is only an approximate relation, it is worthwhile to note that, in this last case, by assuming the scaling $\Dzm_{\perp} \sim \Dzp_{\perp}$, relation (\ref{yaglomrmhd1}) is compatible with the scaling law $\Dzp_{\perp} \sim r_\perp^{2/3}$, which imediately leads to the Kolmogorov spectrum $E(k_{\perp}) \sim k_{\perp}^{5/3}$ predicted for anisotropic MHD turbulence \cite{anisotropia}.

\section{Conclusions}

To conclude, we reviewed the derivation of a general Yaglom's equation for MHD turbulence where two fields are coupled. When this equation is satisfied a turbulent cascade is at work. The most general equation (\ref{anisotropa}) that relates the third-order mixed tensor to the dissipation rate tensor, is valid in the anisotropic and nonhomogeneous case. By using homogeneity, we derive the equation (\ref{anisotropa1}), that is valid in presence of anisotropy, while to derive the usual Yaglom's law (\ref{yaglom}) the local isotropy assumption is required. Moreover, the tensor containing the pressure has zero diagonal elements, but is completely zero only when local isotropy is assumed. As far as the local isotropy assumption is considered, while for usual fluid flows the return to isotropy at small-scales is assured, even if they are anisotropic at large-scale~\cite{returntoisotropy}, in general th MHD flows cannot return completely to isotropy at small scales~\cite{carboveltri,anisotropy1,anisotropy2}. This in turn means that if we want to fully investigate the turbulent cascades in anisotropic MHD flows, the off-diagonal elements of the third-order mixed tensor cannot be disregarded. Of course the tensor term related to pressure cannot be eliminated, while if we cannot assume local isotropy the relation (\ref{anisotropa1}) cannot reduces to the usual Yaglom's law obtained in Ref.s~\cite{pp} and investigated experimentally in Ref.s~\cite{noi,smith}. Of course, since the tensor term $\Pi_{ij}$ cannot be calculated using solar wind data, we cannot have any feeling of the relative importance of this term and the term containing the dissipation rate tensor in anisotropic MHD turbulence. High-resolution numerical simulations for anisotropic MHD turbulence should be used as a first approach.

\acknowledgments{We acknowledge useful discussions with W.H.~Matthaeus.}

\end{document}